\documentclass{lmcs} 
\pdfoutput=1
\usepackage[utf8]{inputenc}

\usepackage{lastpage}
\lmcsdoi{22}{1}{12}
\lmcsheading{}{\pageref{LastPage}}{}{}%
{Aug.~13,~2024}{Feb.~16,~2026}{}


\usepackage{hyperref}

\usepackage{graphicx}
\usepackage{subcaption}

\usepackage{booktabs}
\usepackage{tabularx}
\newcolumntype{C}{>{\centering\arraybackslash}X}
\usepackage{array}
\usepackage{ragged2e}

\usepackage{tikz}
\usetikzlibrary{arrows}
\usetikzlibrary {arrows.meta}

\usepackage{stmaryrd}
\usepackage{amssymb}
\usepackage{wasysym}

\usepackage{amscd}
\usepackage{amstext}
\usepackage{amsthm}
\usepackage{amsmath}

\usepackage{xspace}

\newcommand{\bool}[0]{{\texttt{\scriptsize bool}}}
\newcommand{\nset}[0]{{\texttt{\scriptsize nset}}}
\newcommand{\qset}[0]{{\texttt{\scriptsize qset}}}
\newcommand{\true}[0]{{\mathtt{T}}}
\newcommand{\false}[0]{{\mathtt{F}}}

\newcommand{\LTL}[0]{\textsf{LTL}\xspace}
\newcommand{\MTL}[0]{\textsf{MTL}\xspace}
\newcommand{\PastLTL}[0]{\textsf{PastLTL}\xspace}

\newcommand{\PastMTL}[0]{\textsf{PastMTL}\xspace}

\DeclareMathOperator{\since}{\texttt{S}}
\DeclareMathOperator{\once}{\texttt{P}}
\DeclareMathOperator{\hist}{\texttt{H}}
\DeclareMathOperator*{\prev}{\texttt{Y}}

\DeclareMathOperator{\dom}{\text{dom}}

\newcommand{\V}[0]{\mathbf{V}} 
\newcommand{\X}[0]{\mathbf{X}} 
\newcommand{\Y}[0]{\mathbf{Y}} 
\newcommand{\y}[0]{y}          

\newcommand{\xt}[2]{\texttt{[{\tiny#1:#2}]}}
\newcommand{\xqt}[2]{\texttt{[{\tiny#1:#2}]}}

\newcommand{\histx}[2]{\hist_{\xt{#1}{#2}}}
\newcommand{\oncex}[2]{\once_{\xt{#1}{#2}}}
\newcommand{\sincex}[2]{\since_{\xt{#1}{#2}}}

\newcommand{\sincexq}[2]{\since_{\xqt{#1}{#2}}}

\begin{document}

\title[Online Monitoring of Metric Temporal Logic]{Online Monitoring of Metric Temporal Logic\\using Sequential Networks}

\author[Dogan Ulus]{Dogan Ulus\lmcsorcid{0000-0002-5090-1769}}[]

\address{Boğaziçi University, Istanbul, Türkiye}	
\email{dogan.ulus@bogazici.edu.tr}  




\begin{abstract}
Metric Temporal Logic (\MTL) is a popular formalism to specify temporal patterns with timing constraints over the behavior of cyber-physical systems with application areas ranging in property-based testing, robotics, optimization, and learning.
This paper focuses on the unified construction of sequential networks from \MTL specifications over discrete and dense time behaviors to provide an efficient and scalable online monitoring framework.
Our core technique, future temporal marking, utilizes interval-based symbolic representations of future discrete and dense timelines.
Building upon this, we develop efficient update and output functions for sequential network nodes for timed temporal operations.
Finally, we extensively test and compare our proposed technique with existing approaches and runtime verification tools.
Results highlight the performance and scalability advantages of our monitoring approach and sequential networks.

\end{abstract}

\maketitle

\section{Introduction}
\label{sec:intro}
Monitoring temporal behaviors of complex engineered systems during their execution has important application areas ranging from system verification and anomaly detection to supervisory control.
As modern computing systems grow increasingly complex, requiring highly interactive and sophisticated features, the need for effective and more versatile monitoring solutions increases.
Such high levels of complexity necessitate runtime monitoring of a vast array of temporal properties, including timing constraints and adherence to well-defined behavioral patterns.
Beyond functionality, the concern for performance is also crucial in any runtime monitoring activity, given the inherent overhead introduced to these systems. 
Therefore, fast and versatile runtime monitoring solutions are important assets for ensuring the correctness and smooth operation of these complex systems.

It is often desirable to construct efficient runtime monitors automatically from high-level declarative specifications that describe the system behavior in an unambiguous language.
Initially proposed for formal verification, Linear-time Temporal Logic (\LTL)~\cite{pnueli1977temporal} and its timed extensions, such as Metric Temporal Logic (\MTL)~\cite{koy90}, have become popular formalisms in academia and industry to specify the temporal behavior of real-time reactive systems.
These formalisms have found diverse application areas in robotics, optimization, and property-based testing~\cite{bartocci2018specification, sanchez2019survey}.

This work presents a unified approach for constructing runtime monitors over discrete and dense time behaviors from \LTL and \MTL specifications. 
Leveraging algebraic sequential networks, our proposed approach simplifies the transition from untimed to timed specifications and achieves efficient and scalable runtime monitor construction across both time models. 
We study and emphasize the structural advantages of sequential networks over finite automata, another general solution for runtime monitor construction.
Unlike automata-based constructions, sequential networks offer substantial benefits in compositionality, extensibility, implementability, and scalability, all while preserving functional equivalence in Boolean contexts. 
Consequently, these structural advantages position sequential networks as a powerful and adaptable model of computation for tackling diverse runtime monitoring tasks within complex real-time systems.

Sequential networks in this paper are directly constructed from the past fragments of \LTL and \MTL.
The restriction to the past temporal connectives is twofold: 
(1) Future-oriented (acausal) monitoring is inherently more expensive than past-oriented (causal) monitoring.
The worst-case exponential cost of bookkeeping, among all possibilities in the future, cannot be avoided unless the output at time $t$ is delayed by some duration $d$ depending on the formula.
(2) However, the practical value of delaying seems nonexistent for a truly online/reactive setting as we need an output from the monitor at the current time $t$ rather than the time $t+d$, which may be too late.
This is especially important when the monitor's output is used to make a timely decision, as in the supervisory/reactive control systems.
Consequently, we consider future temporal operators to be a costly feature that does not offer significant practical benefits in online monitoring applications, and we restrict ourselves to the past fragment for online monitoring applications in this paper. 
The basic technique behind constructing sequential networks from past temporal logic formulas involves associating each subformula with a state variable to store relevant information and subsequently updating/manipulating them at each time step.
This paper introduces the \emph{future temporal marking technique} as a novel and effective approach to handling timed operators for sequential network constructions from \PastMTL over discrete and dense time temporal behaviors.
We propose a novel set of update and output equations for each timed operator to mark future time intervals according to timing constraints in the formula.
Notably, our dense time construction seamlessly extends the discrete time construction, avoiding the need for a complete overhaul or reliance on naive discretization as in earlier attempts. 
This facilitates smooth transitions between untimed and timed specifications, as well as between discrete and dense time models, paving the way for a unified treatment of temporal operators across diverse applications and use cases. 
As a result, our approach enables the construction of simpler and more extensible monitors, leading to improved efficiency and adaptability for a wider range of temporal logic monitoring tasks.

The structure of the paper is as follows.
Section~\ref{sec:pre} is dedicated to definitions of sequential networks and temporal logic over discrete and dense time behaviors.
Section~\ref{sec:discrete} and Section~\ref{sec:dense} present and explain our discrete time and dense time sequential network constructions from \PastMTL specifications.
Section~\ref{sec:eval} presents our online discrete and dense time monitoring framework, Reelay\footnote{\url{https://github.com/doganulus/reelay}}, as well as the evaluation and benchmark results.
We then summarize related work~in Section~\ref{sec:related} before concluding with a summary of our contributions and promising avenues for future research in Section~\ref{sec:conc}.

\section{Preliminary Definitions and Background}
\label{sec:pre}

In this paper, we use both discrete and dense time domains, denoted as $\mathbb{T}$ in general.
Specifically, we consider the set $\mathbb{N}$ of non-negative integers and the set $\mathbb{Q}_{>0}$ of positive rational numbers as our discrete and dense time domains, respectively.

\subsection*{Sequential Networks}
\label{sec:sequential}
The online monitoring task can be naturally described as a sequence transformation~\cite{raney1958sequential} from temporal behaviors to monitor verdicts for each time step.
A sequential network is an abstract machine that consists of a finite set $\mathcal{C}$ of computation nodes and implements a sequence-to-sequence transformation, which yields an output sequence $\y_{1}\y_{2}\dots \y_{k}\dots$ from an input sequence $\X_{1}\X_{2}\dots \X_{k}\dots$ given. 
Each node $c\in\mathcal{C}$ is associated with an update function $V_c$ and an output function $Y_c$.
These node functions contribute to the network’s overall update function $V$ and output function $Y$.
The state valuation vector $\V_{k}$ collectively represents the internal state of network nodes at time step $k$, and $\V_{0}$ is called the initial state valuation vector. 
At each time step $k$, the network updates its state vector $\V_k$ by calling node update functions and yielding the current output value $\y_k$ with respect to the previous state vector $\V_{k-1}$ and the current input vector $\X_k$.
We then completely characterize a sequential network by the following elements:
\begin{enumerate}
\item The initial state valuation vector $\V_0$ of length $|\mathcal{C}|$
\item The node update functions $V_c: \V_{k}(c) \Longleftarrow V_{c}(\V_{k-1},\ \X_{k})$ for each node $c\in\mathcal{C}$ 
\item The network output function $Y: \Y_{k} \Longleftarrow Y(\V_{k-1},\ \X_{k})$
\end{enumerate}

\noindent Data types of inputs, outputs, and state variables of the network may vary across applications. 
The simplest class of sequential networks, in which all data types are Booleans, is called Boolean sequential networks or digital sequential circuits when realized using Boolean logic gates and memory elements.
The class of Boolean sequential networks precisely recognizes regular languages and is thus functionally equivalent to finite automata.
The term algebraic sequential networks denotes generalizations of Boolean sequential networks into other compatible algebraic structures, such as the algebra of sets.

\subsection*{Past Linear-time Temporal Logic}

Past Linear-time Temporal Logic (\PastLTL) extends the propositional logic with past temporal modalities such as the \emph{Previously} ($\prev$), \emph{past Always} ($\hist$), \emph{past Eventually} ($\once$), and \emph{Since} ($\since$). \PastLTL provides a robust framework for expressing and reasoning about temporal ordering between system states.

Given a finite set $P$ of atomic predicates, the formulas of \PastLTL are inductively built using the following grammar:
\begin{equation*}
\varphi\ \coloneqq\ \top\ |\  \bot\ |\ p \ |\ \neg \varphi\ |\ \varphi_{1} \wedge \varphi_{2}\ |\ \varphi_{1} \since \varphi_{2}
\end{equation*}
where $p \in P$.
The truth of an arbitrary \PastLTL formula $\varphi$ at a given time instant $t$ over an arbitrary discrete time behavior $w$, denoted as $(w, t) \vDash \varphi$, is defined inductively in the following.
First, the following statements define the propositional fragment of \PastLTL.
\begin{equation}
\begin{array}{rlcl}
(w,t) &\vDash \top & \ \leftrightarrow\ & \true\\
(w,t) &\vDash \bot & \ \leftrightarrow\ & \false\\
(w,t) &\vDash p & \ \leftrightarrow\ & w_p(t) = \true  \\
(w,t) &\vDash \neg \varphi & \ \leftrightarrow\ & (w,t) \nvDash \varphi \\
(w,t) &\vDash \varphi_{1} \wedge \varphi_{2} & \ \leftrightarrow\ & (w,t) \vDash \varphi_{1} \text{ and } (w, t) \vDash \varphi_{2}
\end{array}
\label{DefProp}
\end{equation}
It is clear how to extend the definitions of negation ($\neg$) and conjunction ($\wedge$) of the other Boolean operators of disjunction ($\vee$), implication ($\rightarrow$), double implication ($\leftrightarrow$), etc.

In the literature, temporal modalities can have reflexive and irreflexive versions, which differ in how they handle the current time point. The distinction between these two versions becomes significant when discussing the expressive power. First, the following statement defines the irreflexive version of the Since modality:
\begin{equation}
\begin{array}{rlcl}
(w,t) &\vDash \varphi_{1}\since \varphi_{2} & \ 
 \leftrightarrow\ & \exists t' < t.\ (w, t') \vDash \varphi_{2}\quad \text{and}\quad  \forall t' < t'' < t.\ (w, t'') \vDash \varphi_{1}
\end{array}
\label{DefS}
\end{equation}
Other commonly used past temporal modalities \emph{Previously} ($\prev$), \emph{past Eventually} ($\once$), and \emph{past Always} ($\hist$) can be derived from the \emph{Since} modality using equivalences 
\begin{alignat}{5}
(a)\ \prev\varphi &\equiv  \bot\since \varphi,  &\quad\quad&
(b)\ \once\varphi &\equiv\top\since \varphi,    &\quad\text{ and }\quad&
(c)\ \hist\varphi &\equiv \neg\once\neg\varphi.  &&\label{EqLTL}
\end{alignat}
\noindent Then we can add the following statements to the satisfaction relation:
\begin{alignat}{2}
(w,t) &\vDash \prev\varphi & \ \leftrightarrow\ & (w,\ t - 1) \vDash \varphi \label{DefY}\\
(w,t) &\vDash \once\varphi & \ \leftrightarrow\ & \exists t' < t.\ (w,t') \vDash \varphi\label{DefP}\\
(w,t) &\vDash \hist\varphi & \ \leftrightarrow\ & \forall t' < t.\ (w,t') \vDash \varphi\label{DefH}
\end{alignat}
Observe that the irreflexive versions of temporal modalities exclude the current time point~$t$ from their range of quantification. On the other hand, the reflexive versions differ from their irreflexive counterparts by additionally considering the current time point. The following statement defines the reflexive version of the Since modality:
\begin{equation}
\begin{array}{rlcl}
(w,t) &\vDash \varphi_{1} \since' \varphi_{2} & \ 
 \leftrightarrow\ & \exists t' \leq t.\ (w, t') \vDash \varphi_{2}\quad \text{and}\quad  \forall t' < t'' \leq t.\ (w, t'') \vDash \varphi_{1}
\end{array}
\label{DefSx}
\end{equation}
From this definition, we can derive reflexive versions of the Past Eventually and the Past Always operators similarly:
\begin{alignat}{2}
(w,t) &\vDash \once'\varphi & \ \leftrightarrow\ & \exists t' \leq t.\ (w,t') \vDash \varphi\label{DefPx}\\
(w,t) &\vDash \hist'\varphi & \ \leftrightarrow\ & \forall t' \leq t.\ (w,t') \vDash \varphi\label{DefHx}
\end{alignat}
Yet it is well known that the reflexive version of the \emph{Since} modality is expressively weaker than the irreflexive one as it cannot express the \emph{Previously} modality using Equivalence~\ref{EqLTL}(a) or another way. Therefore, the \emph{Previously} modality must be added explicitly to the grammar when using reflexive definitions to preserve expressiveness.

\subsection*{Past Metric Temporal Logic}
Past Metric Temporal Logic (\PastMTL) extends \PastLTL with a timed variant of the \emph{Since} modality, denoted by $\sincex{a}{b}$, where \texttt{a} and \texttt{b} denote lower and upper bounds restricting the range of quantification for the modality.
We omit $\texttt{a}$ and $\texttt{b}$ in the notation if there is no constraint on the lower and upper end, respectively.
%
%
We refer to a temporal modality untimed if there is no constraint on both ends.
The following statement defines the timed and irreflexive version of the \emph{Since} modality:
\begin{equation}
\begin{array}{rlcl}
(w, t) &\vDash \varphi_{1}\sincex{a}{b} \varphi_{2} & \ 
 \leftrightarrow\ & \exists t' < t.\ (w, t') \vDash \varphi_{2} \text{ and }\\
 & & & \quad\forall t' < t'' < t.\ (w, t'') \vDash \varphi_{1}\text{ and }\\
 & & & \quad\quad t-b \leq t' < t-a\\
\end{array}
\label{DefSd}
\end{equation}
From this definition, timed versions of \emph{Past Eventually} and \emph{Past Always} operators are derived using Equivalences~\ref{EqLTL}(b) and~\ref{EqLTL}(c), respectively, as follows.
\begin{alignat}{2}
(w,t) &\vDash \oncex{a}{b}\varphi & \ \leftrightarrow\ & \exists t' < t.\ (w,t') \vDash \varphi\text{ and } t-b \leq t' < t-a \label{DefPd}\\
(w,t) &\vDash \histx{a}{b}\varphi & \ \leftrightarrow\ & \forall t' < t.\ (w,t') \vDash \varphi\text{ and } t-b \leq t' < t-a \label{DefHd}
\end{alignat}
Notice that the formula $\bot\since \varphi$ is equivalent to $\bot$ for any formula $\varphi$ under dense time interpretation, unlike the discrete case. 
Hence it is meaningless to define the \emph{Previously} modality as one should expect in dense time.

However, when we interpret \MTL over discrete time behaviors, we can always replace timed irreflexive \emph{Since} connective with the \emph{Previously} connective as defined in Equation~\ref{DefY} and timed reflexive \emph{Since} defined as follows:

\begin{equation}
\begin{array}{rlcl}
(w, t) &\vDash \varphi_{1}\sincex{a}{b}' \varphi_{2} & \ 
 \leftrightarrow\ & \exists t' \leq t.\ (w, t') \vDash \varphi_{2} \text{ and }\\
 & & & \quad\forall t' < t'' \leq t.\ (w, t'') \vDash \varphi_{1}\text{ and }\\
 & & & \quad\quad t-b \leq t' \leq t-a\\
\end{array}
\label{DefSdx}
\end{equation}
From this definition, we can derive reflexive versions of the timed Past Eventually and the timed Past Always operators similarly:
\begin{alignat}{2}
(w,t) &\vDash \oncex{a}{b}'\varphi & \ \leftrightarrow\ & \exists t' \leq t.\ (w,t') \vDash \varphi\text{ and } t-b \leq t' \leq t-a \label{DefPdx}\\
(w,t) &\vDash \histx{a}{b}'\varphi & \ \leftrightarrow\ & \forall t' \leq t.\ (w,t') \vDash \varphi\text{ and } t-b \leq t' \leq t-a \label{DefHdx}
\end{alignat}

This paper adopts a careful and consistent approach to applying reflexive and irreflexive semantics of temporal modalities. 
Specifically, we exclusively use reflexive semantics (Eq.~\ref{DefY}, \ref{DefSx} and~\ref{DefSdx}) of temporal modalities when interpreting formulas over discrete time behaviors. 
On the other hand, we exclusively use irreflexive semantics (Eq.~\ref{DefS} and~\ref{DefSd}) when interpreting formulas over dense time behaviors. 
In this manner, our objective is to establish well-behaving default presets for online temporal logic monitoring applications.
Once this is understood, we drop the prime notation for reflexive modalities and use the same notation for temporal modalities across discrete and dense time settings.

\section{Discrete Time Sequential Network Constructions}
\label{sec:discrete}

This section describes how to construct sequential networks from \PastLTL and \PastMTL specifications for discrete time behaviors. We assume that we observe a finite set of propositions at each discrete time point, which can be represented as a Boolean vector. 
The sequential network is fed with these vectors incrementally, one at a time. To track the passage of time for timed properties, we use a global time counter that increments with each discrete time step, corresponding to the sequence index.
Finally, it's important to remember that in the discrete setting, we always use reflexive semantics for temporal modalities and specify timing constraints over integers.

\subsection{Sequential Networks from PastLTL}
\label{sec:ltl2seq}

The \PastLTL monitor construction presented here serves as a crucial preparation step for constructing monitors from timed specifications in the following sections. 
This construction shares significant similarities with the dynamic programming technique described in ~\cite{havelund2004monitoring} while seemingly distinct in terminology and presentation. 
Yet these differences and changing the point of view are important in extending the approach toward metric extensions.

Given a \PastLTL formula, we construct a sequential network consisting of nodes representing each subformula of the formula. 
Each node maintains a Boolean state variable, collectively forming a state valuation vector $\V$.
Initially, all nodes are set to false, such that $\V_0=\bot$.
At each time point $k$, the network's update equations determine the new state valuation of each node based on its operator type and the current valuations of other nodes.
These update equations defined based on the operator type are as follows:
\begin{equation}
\begin{array}{rcl}
\V_{k}(p) & \Longleftarrow & \X_{k}(p) \coloneqq \begin{cases} \true & \text{if the proposition } p \text{ holds at time point } k\\ \false & \text{otherwise.}\end{cases}\\
\V_{k}(\neg\varphi) & \Longleftarrow & \neg \Y_{k}(\varphi)\\
\V_{k}(\varphi_1 \wedge \varphi_2) & \Longleftarrow & \Y_k(\varphi_1) \wedge \Y_k(\varphi_2)\\
\V_{k}(\varphi_1 \vee \varphi_2) & \Longleftarrow & \Y_k(\varphi_1) \vee \Y_k(\varphi_2)\\
\V_{k}(\prev \varphi) & \Longleftarrow & \Y_{k-1}(\varphi)\\
\V_{k}(\once \varphi) & \Longleftarrow & \Y_{k}(\varphi) \vee \V_{k-1}(\once \varphi)\\
\V_{k}(\hist \varphi) & \Longleftarrow & \Y_{k}(\varphi) \wedge \V_{k-1}(\hist \varphi)\\
\V_{k}(\varphi_1\since \varphi_2) & \Longleftarrow & \Y_k(\varphi_2)\ \vee\ \big(\Y_{k}(\varphi_1) \wedge\ \V_{k-1}(\varphi_1\since \varphi_2)\big)
\end{array}
\label{eq:pastltl_update}
\end{equation}
where $\X_{k}$ is the propositional input vector. 
The output function $\Y_{k}(\varphi) \Longleftarrow \V_{k}(\varphi)$ determines the output at the time point $k$. 
Since the output function is trivial for the untimed case, it is ignored in monitor constructions from untimed specifications like~\cite{havelund2004monitoring,monitor-bdd}.
However, we must distinguish the state and output values of network nodes when dealing with timed extensions in the following.

We now illustrate the compositionality of sequential network construction with an example. 
Consider a \PastLTL formula $\varphi \coloneqq (p \vee q)\since \neg r$, which contains three propositions and three non-leaf subformulas. 
We then construct a sequential network from $\varphi$, which has six nodes (all initialized to false) with update equations and the output function given in Table~\ref{tab:network-ltl}.
\begin{table}[b]
\begin{equation*}
    \begin{array}{rcl}
    \toprule
    \V_{k}(p):\bool & \Longleftarrow & \X_{k}(p):\bool\\
    \V_{k}(q):\bool & \Longleftarrow & \X_{k}(q):\bool\\
    \V_{k}(r):\bool & \Longleftarrow & \X_{k}(r):\bool\\
    \V_{k}(\neg r):\bool & \Longleftarrow & \neg \Y_{k}(r):\bool\\
    \V_{k}(p \vee q):\bool & \Longleftarrow & \Y_{k}(p):\bool\ \vee\ \Y_{k}(q):\bool\\
    \V_{k}(\varphi):\bool & \Longleftarrow & \Y_{k}(\neg r):\bool\ \vee\ \Big(\Y_{k}(p \vee q):\bool \wedge \V_{k-1}(\varphi):\bool\Big)\\
    \midrule
    \Y_k(\varphi):\bool & \Longleftarrow & \V_{k}(\varphi):\bool\\
    \bottomrule
    \end{array}
\end{equation*}
\caption{Sequential network constructed from the \LTL formula \mbox{$\varphi \coloneqq (p \vee q)\since \neg r$}. Equations are annotated by type information for the state and input variables, which are always Boolean for networks constructed from \PastLTL specifications.}
\label{tab:network-ltl}
\end{table}
Easily seen, we can construct the same sequential network from previously constructed monitors of $(p \vee q)$ and $\neg r$.
Given two monitors for \PastLTL formulas $\varphi_1 \coloneqq (p \vee q)$ and $\varphi_2 \coloneqq \neg r$, we can obtain a new monitor for $\varphi_1\since \varphi_2$ by joining state variables from both monitors plus adding a new state variable for the (topmost) Since operator.
Then, the outputs of $\varphi_1$ and $\varphi_2$ monitors become arguments of the update equation of the new state.
In other words, output equations of $\varphi_1$ and $\varphi_2$ are embedded into the update equation of $\varphi_1\since \varphi_2$ at the composition.
Note that the update order for state variables is critical as we use reflexive semantics.
Therefore, the implementation must ensure subformulas are updated earlier than their parents.
This is always possible for \LTL formulas due to the acyclic nature of \LTL parse trees, which allows for methods like post-order traversal or topological sorting to establish a well-defined evaluation order.

\subsection{Sequential Networks from PastMTL}
\label{sec:mtl2seq}
This section extends the \PastLTL monitor construction in Section~\ref{sec:ltl2seq} towards \PastMTL specifications.
Timed specifications differ from untimed specifications in that temporal distances between events and states play a role in determining satisfaction.
Suppose we want to evaluate a \PastMTL formula $\psi_1\sincex{a}{b}\psi_2$ at the current time point~$k$.
Without timing constraints, we need to check the past time points where $\psi_2$ holds and ensure $\psi_2$ holds since then. 
This is insufficient for the timed specification as we also need to check the distance between $k$ and the past time points where $\psi_{2}$ holds.
The naive way to address this problem is formula discretization, that is, encoding timing constraints as a series of the Previous operator.
However, this technique does not scale for large timing constraints, and the monitoring performance is heavily degraded while the timing constraints are getting larger.
A better solution for the problem is to keep the history bounded by the window $[k-b, k]$ and check timing constraints retroactively.
Although this technique performs better than naive discretization, it still suffers from large timing constraints when the number of events increases in the window.
In Section~\ref{sec:eval-part1}, we present our performance experiments using existing \MTL monitoring tools using these techniques~\cite{monpoly,monpoly2,aerial}.

In this paper, we propose an alternative technique, called \emph{future temporal marking}, which labels future temporal points with the corresponding information to be used when the future time arrives.
For example, consider online monitoring of the \PastMTL formula $\varphi\coloneqq\oncex{a}{b}\psi$.
Unlike existing techniques that rely on keeping a bounded history of output values for the formula $\psi$, our technique proactively marks all future time points in $[k+a, k+b]$ whenever the formula $\psi$ holds at time point~$k$.
At a future time point $k'$, we know the formula $\varphi$ needs to be evaluated to true if $k'$ is already marked, false otherwise.
Such a proactive approach necessitates maintaining a subset of future time points marked for each timed operator as time progresses.
This observation is our starting point for constructing a discrete time sequential network that can manipulate sets of integers.

Following the \PastLTL construction in Section~\ref{sec:ltl2seq}, we begin our construction by associating each propositional and untimed subformula with a Boolean state variable and updating them using the same set of update equations in Equation~\ref{eq:pastltl_update}.
Therefore, the construction for the untimed fragment remains unaltered in the discrete setting.
For the timed fragment, we create nodes for each timed operator $\varphi$ that maintains an integer-set (\nset) valued state variable $\V_{k}(\varphi)\subseteq [k, \infty)$.
This integer set essentially represents a subset of future integer time points, and this property is ensured by intersecting $\V_{k}(\varphi)$ with $[k, \infty)$ after an update, an operation referred to as \emph{trimming}.

In the following, we explain and formalize how to update timed state variables for each timed operator.

\paragraph{Timed Past Eventually}
According to the reflexive \PastMTL semantics, the formula $\oncex{a}{b}\psi$ holds at a time point $k$ if the formula $\psi$ holds for some time points in the discrete time interval $[k-b, k-a]$.
For the timed Eventually nodes, we want to mark future time points and maintain the integer-set valued state variable $\V(\oncex{a}{b}\psi)$ to update it at every time point.
This behavior is formally captured in the update equation of timed past Eventually nodes as follows:
\begin{equation}
\centering
\V_k(\oncex{a}{b}\psi):\nset\ \Longleftarrow\
    \begin{cases} 
        \V_{k-1}(\oncex{a}{b}\psi):\nset\ \cup\ [k+a, k+b] & \text{if } \Y_k(\psi):\bool\\ 
        \V_{k-1}(\oncex{a}{b}\psi):\nset & \text{otherwise.}
    \end{cases}
\end{equation}
where the Boolean-valued function $\Y_k(\psi)$ denotes the output equation of the formula~$\psi$ and the initial state $\V_{0}$ is defined to be empty.
Here, the update equation marks future time points based on the subformula's truth value and timing constraints, while previously marked time points remain marked.
It is easy to see that the current time point $k$ must be in $\V_k(\oncex{a}{b}\psi)$ by definition if the formula $\oncex{a}{b}\psi$ holds at~$k$.
From this observation, we define the output function $\Y_k(\oncex{a}{b}\psi)$ to be a membership test as follows:
\begin{equation}
\centering
\Y_k(\oncex{a}{b}\psi):\bool\ \Longleftarrow\ k \in \V_{k}(\oncex{a}{b}\psi)
\end{equation}
\noindent Notice that unlike the untimed case, where both state valuations and output values are Boolean, the discrete timed case introduces different data types for valuations and output values. 
We consider this fact to be the general case for sequential networks and regard the untimed case to be a specialization of the general framework established in this paper.

\begin{table}[b]
\begin{equation*}
\centering
\setlength{\arraycolsep}{6pt}
\begin{array}{rcl}
\toprule
\V_{k}(p):\bool & \Longleftarrow & \X_{k}(p):\bool\\
\V_{k}(q):\bool & \Longleftarrow & \X_{k}(q):\bool\\
\V_{k}(\psi_{1}):\bool & \Longleftarrow & \Y_{k}(p):\bool\ \vee\ \Y_{k}(q):\bool\\
\V_{k}(\psi_{2}):\nset & \Longleftarrow & \begin{cases} \V_{k-1}(\psi_{2}):\nset\ \cup\ [k+1, k+2] & \text{if } \Y_{k}(\psi_{1}):\bool\\ \V_{k-1}(\psi_{2}):\nset & \text{otherwise.}\end{cases}\\
\V_{k}(\varphi):\nset & \Longleftarrow & \begin{cases} \V_{k-1}(\varphi):\nset\ \cup\ [k+1, k+2] & \text{if } \Y_{k}(\psi_{2}):\bool\\ \V_{k-1}(\varphi):\nset & \text{otherwise.}\end{cases}\\
\midrule
\Y_k(\varphi):\bool & \Longleftarrow & k \in \V_{k}(\varphi):\nset\\
\bottomrule
\end{array}
\end{equation*}
\caption{Sequential network constructed from the formula $\varphi \coloneqq \once_{[1,2]} \once_{[1,2]}(p \vee q)$ where $\psi_{1}\coloneqq p \vee q$ and $\psi_{2}\coloneqq \once_{[1,2]}(\psi_{1})$.}
\label{tab:network-timed-eventuality}
\end{table}

\begin{table}[t]
\begin{equation*}
\setlength{\arraycolsep}{10pt}
\begin{array}{rccccccccc}
\toprule
k && - & 0 & 1 & 2 & 3 & 4 & 5 \\
\midrule
\X(p):\bool &&   & \true & \false & \false & \false & \false & \false \\
\X(q):\bool &&   & \false & \false & \false & \false & \true & \false \\
\midrule
\V(p):\bool && \false & \true & \false & \false & \false & \false & \false \\
\V(q):\bool && \false & \false & \false & \false & \false & \true & \false \\
\V(\psi_{1}):\bool&& \false & \true & \false & \false & \false & \true & \false\\
\V(\psi_{2}):\nset&& \emptyset & [1,2] & [1,2] & \{2\} & \emptyset & [5,6] & [5,6]\\
\V(\varphi):\nset&& \emptyset & \emptyset & [2,3] & [2,4] & [3,4] & \{4\} & [6,7]\\
\midrule
\Y(\varphi):\bool &&  & \false & \false & \true & \true & \true & \false\\
\bottomrule
\end{array}
\end{equation*}
\caption{An example execution of the sequential network in Table~\ref{tab:network-timed-eventuality}.}
\label{tab:run-timed-eventuality}
\end{table}

In the following, we present an example sequential network construction from the \PastMTL formula $\varphi \coloneqq \oncex{1}{2}\oncex{1}{2}(p \vee q)$, which has three non-leaf subformulas $\psi_{1}: p \vee q$ and $\psi_{2}:\oncex{1}{2}(p \vee q)$ including the formula $\varphi$ itself.
Our construction produces one Boolean and two timed nodes in the resulting network, as shown in Table~\ref{tab:network-timed-eventuality} with corresponding update equations and the output function.
Notice that the output function $k\in \V_{k}(\psi_{2})$ of $\once_{[1,2]}(p \vee q)$ is embedded into the update equation of $\V_{k}(\varphi)$ during the construction.
Table~\ref{tab:run-timed-eventuality} illustrates an example run over a discrete time behavior from the time index $k=0$ to $5$ over the propositions $p$ and $q$ in the first two rows.
The next three rows below denote the valuation of state variables, and the final row denotes the output of the sequential network.
Observe that the timed nodes trim their valuations to be a subset of $[k, \infty)$ as time progresses.
This is important in practice to keep the size of state valuations small.
A minor limitation and our performance experiments related to the size of state valuations are presented in Section~\ref{sec:eval-part1}.

Finally note that the formula $\oncex{1}{2}\oncex{1}{2}(p \vee q)$ is semantically equivalent to a simpler formula of $\oncex{2}{4}(p \vee q)$. 
Therefore, we know it is possible to construct a smaller network to monitor this formula. 
However, in this paper, our constructions follow the formula structure faithfully. 
Although we do not target the most optimal networks, we still recognize the potential for performance improvements through syntactic optimizations such as formula rewriting and common subformula elimination.

\begin{table}[b]
\begin{equation*}
\setlength{\arraycolsep}{10pt}
\begin{array}{rcccccccc}
\toprule
k & - & 0 & 1 & 2 & 3 & 4 & 5 \\
\midrule
\X(p):\bool  &   & \false & \false & \true & \true & \true & \false \\
\midrule
\V(\histx{1}{2} p):\nset  & \emptyset & [1,2] & [1,3] & [2,3] & \{3\} & \emptyset & [6,7]\\
\midrule
\Y(\histx{1}{2} p):\bool &  & \true & \false & \false & \false & \true & \true\\
\bottomrule
\end{array}
\end{equation*}
\caption{An example execution of the sequential network constructed for the formula $\varphi \coloneqq \histx{1}{2} p$ over an discrete time behavior of the proposition $p$.}
\label{tab:run-timed-always}
\end{table}

\paragraph{Timed Past Always} 
According to the reflexive \PastMTL semantics, the formula $\histx{a}{b}\psi$ holds at a time point $k$ if the formula $\psi$ holds for all time points in the discrete time interval $[k-b, k-a]$.
Similar to the timed past Eventually case, we maintain the integer-set valued state variable $\V(\histx{a}{b}\psi)$ for timed past Always nodes and update it at every time point.
To capture this behavior efficiently, we leverage the duality property from Equivalence~\ref{EqLTL}(c) and define the update equation for the timed past Always operator as follows:
\begin{equation}
\V_{k}(\histx{a}{b}):\nset\ \Longleftarrow\ \begin{cases} \V_{k-1}(\histx{a}{b}):\nset\ \cup\ [k+a, k+b] & \text{if } \neg \Y_{k}(\psi):\bool\\ \V_{k-1}(\histx{a}{b}):\nset & \text{otherwise.}\end{cases}
\end{equation}
where the Boolean-valued function $\Y_k(\psi)$ denotes the output equation of the formula~$\psi$.
But, unlike the timed past Eventually case, the update equation now marks new future time points if the subformula does not hold, and we need to check the absence of current time point $k$ in the $\V_{k}(\histx{a}{b})$ to output true. The output function is then defined as:
\begin{equation}
 \Y_k(\histx{a}{b}):\bool\ \Longleftarrow\ k \notin \V_{k}(\histx{a}{b})
\end{equation}
The initial state $\V_{0}(\histx{a}{b})$ is similarly set to the empty set $\emptyset$.
This approach allows the operators $\oncex{a}{b}$ and $\histx{a}{b}$ to seamlessly handle vacuous truth situations where no truth value is assigned before the initial time point.
Table~\ref{tab:run-timed-always} illustrates an example run of the sequential network constructed from the formula $\varphi \coloneqq \histx{1}{2} p$. 

\paragraph{Timed Since} 
According to the reflexive \PastMTL semantics, the formula $\psi_1\sincex{a}{b}\psi_2$ holds at a time point $k$ if $\psi_2$ held at time point~$k'$ in the past between time points~$k-b$ and~$k-a$, and $\psi_1$ has held continuously from~$k'$ to~$k$. 
The timed Since operation is the most general case for our discrete time \PastMTL construction, as the previous cases can be derived from this construction.
Intuitively speaking, online monitoring of timed Since nodes using our approach requires marking future time points whenever the formula $\psi_2$ holds according to timing constraints and removing them when the formula $\psi_1$ ceases to hold.
In other words, timed Since nodes dynamically maintain an integer-set valued state variable based on the output of the formula $\psi_{2}$ and timing constraints.
These integers represent future time points where the timed Since node might be satisfied, while requiring the subformula $\psi_1$ to continuously hold until then.

Based on these observations, we formally define the update equation of the sequential network constructed for the formula $\varphi \coloneqq \psi_{1}\sincex{a}{b}\psi_{2}$ as follows:
\begin{equation}
\V_{k}(\varphi):\nset\ \Longleftarrow\ \begin{cases} 
    \V_{k-1}(\varphi):\nset\ \cup\ [k+a, k+b] & \text{if } \Y_k(\psi_1) \wedge \Y_k(\psi_2)\\ 
    [k+a, k+b] & \text{if } \neg \Y_k(\psi_1) \wedge \Y_k(\psi_2)\\
    \V_{k-1}(\varphi):\nset & \text{if } \Y_k(\psi_1) \wedge \neg \Y_k(\psi_2)\\
    \emptyset & \text{otherwise.}
    \end{cases}
\label{DefUpdSince}
\end{equation}
The initial state $\V_{0}(\psi_1\sincex{a}{b}\psi_2)$ is set to the empty set $\emptyset$ and the output function is similarly defined as a membership test for the timed Since nodes as follows:
\begin{equation}
\Y_k(\psi_1\sincex{a}{b}\psi_2):\bool\ \Longleftarrow\ k \in \V_{k}(\psi_1\sincex{a}{b}\psi_2)
\label{DefSMemCheck}
\end{equation}
These definitions complete our discrete time sequential network construction from the reflexive \PastMTL specifications. 
The irreflexive case is similar to the reflexive. 
The irreflexive update function does not add $k$ to the valuation for the case when $\Y_k(\psi_2)$ holds. 
Therefore, we can obtain the update function for the irreflexive Since operation from Equation~\ref{DefUpdSince} by replacing $[k+a,k+b]$ with $(k+a, k+b]$ when $a=0$ and derive update functions for other irreflexive temporal operators.

Each node in the network synchronously updates its state based on information as defined by their update and output equations.
Table~\ref{tab:run-timed-since} illustrates an example run of the sequential network constructed from the formula $\varphi = p\ \sincex{2}{3} q$. 
At time point $1$, the proposition $q$ is true, and we mark future time points $[3,4]$. 
These potential satisfaction points for the formula will be kept in the valuation set if the proposition $p$ continues to hold, as realized for time points $3$ and $4$. 
The proposition $p$ gets false at time point $5$; therefore, potential satisfaction for time points $6$ and $7$ will not be realized. 
Note that the valuation set has been trimmed at each time step therefore the valuation set at time point 4 does not contain the time point 3. 

\begin{table}[t]
\begin{equation*}
\setlength{\arraycolsep}{10pt}
\begin{array}{rcccccccc}
\toprule
k & - & 0 & 1 & 2 & 3 & 4 & 5 \\
\midrule
\X(p):\bool &   & \false & \false & \true & \true & \true & \false \\
\X(q):\bool &   & \false & \true & \false & \false & \true & \false \\
\midrule
\V(p\ \sincex{2}{3} q):\nset& \emptyset & \emptyset & [3,4] & [3,4] & [3,4] & \{4\} \cup [6,7] & \emptyset\\
\midrule
\Y(p\ \sincex{2}{3} q):\bool &  & \false & \false & \false & \true & \true & \false\\
\bottomrule
\end{array}
\end{equation*}
\caption{An example execution of the sequential network constructed for the formula $\varphi\coloneqq p\ \sincex{2}{3} q$ over discrete time behaviors of $p$ and~$q$.}
\label{tab:run-timed-since}
\end{table}

\medskip
\noindent Finally, we state the correctness of our construction with the following theorem.
\begin{thm}
For any \PastMTL formula $\varphi$ and discrete time behavior $w$, the sequential network output $\Y_k(\varphi)$ evaluates to true at time point $k$ over $w$ iff $(w, k) \vDash \varphi$ holds.
\label{thm:discrete-correctness}
\end{thm}
\begin{proof} We only show the correctness of the timed Since case as the propositional and untimed fragment is straightforward and other temporal operators can be derived from timed Since.
\begin{enumerate}
\item[($\shortrightarrow$)] First, we show that $(w, k) \vDash \psi_{1}\sincex{a}{b}\psi_{2}$ holds if time point $k \in \V_{k}(\psi_{1}\sincex{a}{b}\psi_{2})$. 
Assume that $k \in \V_{k}(\psi_{1}\sincex{a}{b}\psi_{2})$.
By Definition~\ref{DefUpdSince} of the update function, this implies that $k$ was added to the valuation set for some $k^{\dagger} \in [k-b,k-a]$, which satisfies the first and timed condition of the semantic definition in Definition~\ref{DefSdx}.
Furthermore, this implies $\psi_{1}$ holds continuously from the last addition of $k$ until the current time point $k$ as the update function preserves time points from $\V_{k}$ when $\psi_{1}$ holds.
Thus, the semantic definition is satisfied in either case, and we conclude $(w, k) \vDash \psi_{1}\sincex{a}{b}\psi_{2}$.

\item[($\shortleftarrow$)] Second, we show that $(w, k) \nvDash \psi_{1} \sincex{a}{b} \psi_{2}$ holds if the time point~$k \notin \V_{k}(\psi_{1}\sincex{a}{b}\psi_{2})$.
Assume that  $k \notin \V_{k}(\psi_{1}\sincex{a}{b}\psi_{2})$.
There are two possibilities:
\begin{enumerate}[(i)]
\item The time point $k$ is never added to $\V_{k}$. Then, there exists no time point $k^{\dagger} \in [k-b,k-a]$ that satisfies $\psi_{2}$, thus $(w, k) \nvDash \psi_{1}\sincex{a}{b}\psi_{2}$.
\item The time point $k$ is initially added to $\V_{k}$ at some time point $k^{\dagger} \in [k-b,k-a]$ and removed at a later step. This means there exists a time point $k^{\ddagger}$ such that $k^{\dagger} < k^{\ddagger} \leq k$ where $\psi_1$ does not hold as the update function only removes time points from $\V_{k}$ when $\psi_{1}$ does not hold. Thus, $(w, k) \nvDash \psi_{1}\sincex{a}{b}\psi_{2}$.
\end{enumerate}
The semantic definition is violated in either case, and we conclude $(w, k) \nvDash \psi_{1}\sincex{a}{b}\psi_{2}$.
\end{enumerate}
This completes both directions of the equivalence proof, establishing the correctness of the update function for the timed Since operator and our construction.
\end{proof}

\section{Dense Time Sequential Network Construction}
\label{sec:dense}

This section explains sequential network constructions over the dense time behaviors from \PastMTL specifications, extending our discrete time constructions in the previous section.

The first difference between discrete and dense time settings is the representation of temporal behaviors. 
Discrete time behaviors are naturally described as a finite sequence of atomic observations indexed by discrete time points.
Such finite representation is not straightforward for dense time behaviors as we lack a standard successor relation for dense domains.
One solution is timed event sequences that sample a finite subset of time points from the dense time domain and order them by the standard ordering relation.
However, timed event sequences have a severe limitation in that the state of monitors can be updated only at sample points, and a non-sufficiently sampled time domain may cause the monitor to yield a wrong verdict~\cite{reynolds2016metric,basin2018algorithms}.

An alternative solution is to represent dense time behaviors as finite sequences of time intervals~\cite{realtime-logic}, that is, non-empty convex subsets of the time domain. 
Although the traditional continuum representation is more natural and does not have the aforementioned irregularities of timed event sequences, this representation is much less developed for dense time monitoring algorithms~\cite{reynolds2016metric,basin2018algorithms}. 
One possible reason is that any algorithm working on the traditional continuum must deal with all the challenges of real analysis, such as boundary inclusion/exclusion, limit conditions, epsilon neighborhoods, and Zeno paradoxes.
However, these mathematical artifacts hardly matter in monitoring and lead to unnecessarily complicated algorithms.
Practical monitor implementations thus restrict dense time behaviors to have finite variability ---meaning a finite number of discontinuities exists within any bounded interval--- and employ a single interval type in practice.
These practices are departures from the initial assumptions of the continuum, but nonetheless, they can be formalized using a restriction to finite representation and a coarser topology on the time domain.
Yet a deeper discussion is beyond our scope in this paper.

This paper follows the latter solution and represents dense time behaviors as finite sequences of left-open and right-closed intervals.
This particular choice of interval type follows from the fact that the Since operator is left-continuous~\cite{ferrere2019real}, intuitively meaning that the \emph{Since} operation yields and preserves left-open and right-closed interval types on dense domains. 
Therefore, our approach uniformly treats and requires dense time intervals to be left-open and right-closed for our dense time procedures.

We start our dense time monitor construction by assuming we observe a finite set $P$ of propositions. 
We consider a dense time Boolean behavior $w: (t_0, t_n]\to\mathbb{B}^{P}$ is a function that can be represented as a sequence of constant valued segments such that
\begin{equation*}
(t_0, t_1]\to \X_1;\ (t_1, t_2]\to \X_2;\ \dots;\ (t_{n-1}, t_n]\to \X_n
\end{equation*}
where $n\in\mathbb{N}$, $t_0, t_1, \dots, t_n\in \mathbb{Q}$ and $\X_1, \X_2, \dots, \X_n\in \mathbb{B}^{P}$ are propositional Boolean vectors.
For the behavior $w$, the segmentation $\tau(w)$ is defined to be the set of its endpoints $\{t_{0}, t_{1}, t_{2}, \dots, t_{n}\}$.
We use the notation $w_k$ to refer to the $k$-th segment of the behavior $w$, and the function $\dom(k)$ denotes its time domain $(t_{k-1}, t_k]$ for $k=1,2,\dots, n$.
The duration $|w_{k}|$ of the segment $k$ is defined to be $t_{k}-t_{k-1}$.
We avoid zero-duration segments in this formulation; thus, time progress is strict.

We then define quantization and condensation as two common operations on dense time behaviors to help analyze the inherent tradeoff between granularity and efficiency in dense time monitoring.
Quantization partitions a dense time behavior $w$ into an equivalent and evenly segmented behavior $w'$ such that the duration $|w'_k|$ of each segment $k$ is equal within $w'$.
Quantization requires a fixed segment length $\delta$ such that each segment boundary $t$ in the behavior $w$ is an integer multiple of~$\delta$.
The set $\{ k\delta\ |\ k \in\mathbb{N} \}$ is called the base of the quantization generated by $\delta$.
Conversely, condensation merges consecutive segments with the same value, called stuttering segments, into longer segments. 
This process can be applied iteratively until the behavior is maximally condensed, meaning that containing no stuttering segments.
Figure~\ref{fig:two-ops} illustrates the application of quantization and condensation to example dense time behaviors.

\begin{figure}[t]
\centering
\renewcommand\thesubfigure{\roman{subfigure}}
\begin{subfigure}{0.45\linewidth}
\resizebox{\textwidth}{!}{%
\begin{tikzpicture}
\draw[] (0,3) rectangle (2,3.6) node[midway] {A};
\draw[] (2,3) rectangle (5,3.6) node[midway] {B};
\draw[] (5,3) rectangle (8,3.6) node[midway] {A};
\draw[] (8,3) rectangle (10,3.6) node[midway] {B};
\draw[] (10,3) rectangle (12,3.6) node[midway] {C};
\draw [line width=4pt, -{Stealth[length=7mm, round]}] (6,2.5) -- (6,1);
\draw[] (0,0) rectangle (1,0.6) node[midway] {A};
\draw[] (1,0) rectangle (2,0.6) node[midway] {A};
\draw[] (2,0) rectangle (3,0.6) node[midway] {B};
\draw[] (3,0) rectangle (4,0.6) node[midway] {B};
\draw[] (4,0) rectangle (5,0.6) node[midway] {B};
\draw[] (5,0) rectangle (6,0.6) node[midway] {A};
\draw[] (6,0) rectangle (7,0.6) node[midway] {A};
\draw[] (7,0) rectangle (8,0.6) node[midway] {A};
\draw[] (8,0) rectangle (9,0.6) node[midway] {B};
\draw[] (9,0) rectangle (10,0.6) node[midway] {B};
\draw[] (10,0) rectangle (11,0.6) node[midway] {C};
\draw[] (11,0) rectangle (12,0.6) node[midway] {C};
\end{tikzpicture}
}
\caption{Quantization}
\end{subfigure}
\begin{subfigure}{0.05\linewidth}
\phantom{|}
\end{subfigure}
\begin{subfigure}{0.45\linewidth}
\resizebox{\textwidth}{!}{%
\begin{tikzpicture}
\draw[] (0,3) rectangle (1,3.6) node[midway] {A};
\draw[] (1,3) rectangle (2,3.6) node[midway] {A};
\draw[] (2,3) rectangle (3,3.6) node[midway] {B};
\draw[] (3,3) rectangle (4,3.6) node[midway] {B};
\draw[] (4,3) rectangle (5,3.6) node[midway] {B};
\draw[] (5,3) rectangle (6,3.6) node[midway] {A};
\draw[] (6,3) rectangle (7,3.6) node[midway] {A};
\draw[] (7,3) rectangle (8,3.6) node[midway] {A};
\draw[] (8,3) rectangle (9,3.6) node[midway] {B};
\draw[] (9,3) rectangle (10,3.6) node[midway] {B};
\draw[] (10,3) rectangle (11,3.6) node[midway] {C};
\draw[] (11,3) rectangle (12,3.6) node[midway] {C};
\draw [line width=4pt, -{Stealth[length=7mm, round]}] (6,2.5) -- (6,1);       
\draw[] (0,0) rectangle (2,0.6) node[midway] {A};
\draw[] (2,0) rectangle (5,0.6) node[midway] {B};
\draw[] (5,0) rectangle (8,0.6) node[midway] {A};
\draw[] (8,0) rectangle (10,0.6) node[midway] {B};
\draw[] (10,0) rectangle (12,0.6) node[midway] {C};
\end{tikzpicture}
}
\caption{Condensation}
\end{subfigure}
\caption{Quantization and condensation operations side by side}
\label{fig:two-ops}
\end{figure}

Dense-time sequential networks process input behavior incrementally, consuming a single constant-valued segment at each step.
Crucially, the network operates without requiring prior knowledge of the overall segmentation, nor does it impose constraints on segment boundaries, length, or arrival rate.
This flexibility is a key feature of our approach. 
The network handles varying data streams, making it ideal for real-time processing and systems with asynchronous or irregular updates.
While segmentation choices do not affect the network's functionality and output, they may influence monitoring performance considerably. 
Section~\ref{sec:eval-dense} provides a detailed performance analysis, examining how different segmentation schemes influence the performance.
Yet, ultimately, the user or application side retains control over the segmentation.

In the following section, we explain and formalize how to update sequential network nodes for each operator, extending our discrete time constructions from the previous section.

\paragraph{Dense Time Propositional Fragment}
Nodes in our dense time sequential network construction manipulate all current and future time points at a single update step. 
This is similar to the discrete case except that current time points are no longer isolated.
Dense time nodes receive input and produce output for all time points in the current interval.

Dense-time propositional nodes internally represent their state valuation as a subset of the rational numbers, denoted as $\qset$.
For a given dense-time segment $k$, the valuation set $\V_k(p)$ consists of rational time points where the proposition $p$ holds.
This extends to Boolean operations naturally, and we define the update equations for both propositional and Boolean nodes as follows:
\begin{equation}
\begin{array}{rcl}
\V_{k}(p):\qset & \Longleftarrow & \{ t \in \dom(k)\ |\ w_p(t) = \true \}\\
\V_{k}(\neg\varphi):\qset & \Longleftarrow & \dom(k)\ \setminus\ \Y_k(\varphi):\qset\\
\V_{k}(\varphi_1 \wedge \varphi_2):\qset & \Longleftarrow & \Y_k(\varphi_1):\qset\ \cap\ \Y_k(\varphi_2):\qset\\
\V_{k}(\varphi_1 \vee \varphi_2):\qset & \Longleftarrow & \Y_k(\varphi_1):\qset\ \cup\ \Y_k(\varphi_2):\qset\\
\end{array}
\label{eq:pastltl_update_dense}
\end{equation}
where $dom(k) \subseteq\mathbb{Q^{+}}$ denotes the domain of the current segment $k$. 
For the dense time setting, the output of each segment is a subset of its time domain, where membership indicates the formula's validity at that particular point in time.
The output function for propositional nodes is then trivially defined as follows:
\begin{equation}
  \Y_{k}(\varphi):\qset\ \Longleftarrow\ \V_{k}(\varphi):\qset
\end{equation}
Therefore, the network efficiently computes and propagates truth values of dense time points within a segment in a single, unified update.

\paragraph{Dense Timed Since} We handle dense timed Since nodes similar to the discrete time construction with a few modifications for the dense time setting.

First, the network node for timed Since needs to process the output values of their operand nodes over constant segments.
Yet, generally, constant segments in the operand outputs are not synchronized inside the current segment. 
Therefore, we must synchronize dense time output valuations from operands to provide constant-valued local segments before feeding them into our node update procedure.
To this end, we introduce a~secondary local index $l$ that indicates the position of a local segment in the current global segment.
For example, consider that we are processing the $k$-th segment that consists of $L_{k}$ constant-valued local segments over the operand values $\psi_1$ and $\psi_2$ for the formula $\psi_1\sincexq{a}{b}\psi_2$.
The notation $\V_{k,l}$ denotes the valuation of $l$-th constant local segment for both operands in the $k$-th global segment for $l = 1\dots L_{k}$, and similarly $\Y_{k,l}$ denotes its output.

The example in Table~\ref{tab:run-timed-since-dense} illustrates the synchronization operation over dense time behaviors of sub-formulas $\psi_1$ and $\psi_2$.
Dense time behaviors of the operands are sequentially given as four global segments in the example, and we locally synchronize them before local updates of the timed Since operation.
Synchronization may lead to further segmentation on the timeline in which both behaviors hold constant, as shown in the row denoted by  $\psi_1 || \psi_2$.
In the worst case, this leads to $|\Y_{k}(\psi_1)| + |\Y_{k}(\psi_2)| - 1$ local segments inside the global segment $k$.
For this example, the first three global segments are divided into four local segments, and the last is divided into two local segments.

Second, node updates within each global segment are performed through local update steps.
Similar to the discrete time update function in Equation~\ref{DefUpdSince}, the update function for dense timed Since nodes $\varphi = \psi_1\sincexq{a}{b}\psi_2$ for the local segment $(k,l)$ is defined as follows:
\begin{equation}
  \V_{k,l}(\varphi):\qset\ \Longleftarrow\  \begin{cases} 
      \V_{k,\ l-1}(\varphi):\qset\ \cup\ (t+a,\ t'+b] & \text{if } \y_{k,l}(\psi_1)\ \wedge\ \y_{k,l}(\psi_2)\\ 
      (t'+a,\ t'+b] & \text{if } \neg \y_{k,l}(\psi_1)\ \wedge\ \y_{k,l}(\psi_2)\\
      \V_{k,\ l-1}(\varphi):\qset & \text{if } \y_{k,l}(\psi_1)\ \wedge\ \neg \y_{k,l}(\psi_2)\\
      \emptyset & \text{otherwise.}
      \end{cases}
\label{DefUpdSinceDense}
\end{equation}
where $\dom(k,l) = (t, t']$ and $\y_{k,l}$ is the Boolean value for the constant-valued local segment $(k,l)$ for the operands. 
Local updates follow a strict sequential order like global update steps, with the final local step serving as the initial step for the subsequent segment such that $\V_{k, L_{k}} = \V_{k+1, 0}$.
The irreflexive Since operation is left-continuous by the semantic definition in Equation~\ref{DefSd}, meaning that, for all behaviors $w$ and time points $t$; $(w,t)\vDash\varphi$ implies there exists $t'<t$ such that for all $t''\in (t',t). (w,t)\vDash\varphi$.
It is easy to see that the local update function preserves the left-continuity of the valuation set.

Finally, we combine the outputs of local updates to prevent timeline fragmentation at the network interface, which may be problematic when the network's output is used to feed another network in a compositional way.
Analogously to the discrete time output function defined in Equation~\ref{DefSMemCheck}, the output function for a local segment is defined using the set intersection operation as follows:
\begin{equation}
  \Y_{k,l}(\varphi):\qset\ \Longleftarrow\ \V_{k,l}(\varphi):\qset\ \cap\ \dom(k,l):\qset
\end{equation}
This set intersection operation symbolically checks membership for all points within the current time segment and yields all time instants that satisfy the formula for the segment.
The final computation step combines the outputs of individual local steps within each global segment and yields the output of the global segment as a whole. 
This step is achieved through the union operation, effectively preventing any timeline fragmentation that could otherwise arise from the local synchronization approach. 
Leveraging these insights, we formally define the output of the global segment $k$ as a union of the outputs of all $L_{k}$ local segments as follows:
\begin{equation}
  \Y_k(\varphi):\qset\ \Longleftarrow\ \Y_{k,1}(\varphi):\qset\ \cup\ \Y_{k,2}(\varphi):\qset\ \cup\ \dots \cup\ \Y_{k,L_{k}}(\varphi):\qset
\end{equation}
Therefore, a dense time node receives the inputs for the current time interval and yields an output for the same interval as the sequential model of computation dictates.

\begin{table}[t]
\arraycolsep=2pt
\begin{equation*}
\begin{array}{rcccccc}
\toprule
\text{Step}(k): & \phantom{m} & 1 & 2 & 3 & 4\\
\midrule
\text{Domain}:  & &  (0,30] & (30,47] & (47,75] & (75,99]\\
\midrule
\psi_1 :&   & \begin{array}{r}(0,7]\to\false;\\(7,30]\to\true;\end{array} & \begin{array}{r}(30,35]\to\true;\\(35,39]\to\false;\\(39,47]\to\true;\end{array} &\begin{array}{r}(47,49]\to\true;\\(49,63]\to\false;\\(63,75]\to\true;\end{array} & \begin{array}{r}(75,99]\to\true;\end{array} \\
\midrule
\psi_2 :&   & \begin{array}{r}(0,3]\to\false;\\(3,8]\to\true;\\(8,30]\to\false;\end{array} & \begin{array}{r}(30,38]\to\false;\\(38,39]\to\true;\\(39,47]\to\false;\end{array} &\begin{array}{r}(47,70]\to\false;\\(70,75]\to\true;\end{array} & \begin{array}{r}(75,89]\to\true;\\(89,99]\to\false;\end{array} \\
\midrule
\psi_1\, ||\, \psi_2:
& 
& \begin{array}{rcc} & \y_{\psi_1} & \y_{\psi_2}\\(0,3]:& \false & \false\\ (3,7]: &  \false & \true\\ (7,8]: & \true & \true\\ (7,30]: & \true & \false\end{array}            
& \begin{array}{rcc} & \y_{\psi_1} & \y_{\psi_2}\\(30,35]:& \true & \false\\ (35,38]: &  \false & \false\\ (38,39]: & \false & \true\\ (39,47]: & \true & \false\end{array} 
& \begin{array}{rcc} & \y_{\psi_1} & \y_{\psi_2}\\(47,49]:& \true & \false\\ (49,63]: &  \false & \false\\ (63,70]: & \true & \false\\ (70,75]: & \true & \true\end{array} 
& \begin{array}{rcc} & \y_{\psi_1} & \y_{\psi_2}\\(75,89]:& \true & \true\\ (89,99]: &  \true & \false\end{array}\\
\midrule
\V_{k,l}(\varphi) :
&  
& \begin{array}{rc} 1: & \emptyset \\ 2: &  \{(25,31]\}\\ 3: & \{(25,32]\}\\ 4: & \{(25,32]\}\end{array} 
& \begin{array}{rc} 1: & \{(30,32]\}\\ 2: &  \emptyset \\ 3: & \{(57,63]\}\\ 4: & \{(57,63]\}\end{array} 
& \begin{array}{rc} 1: & \emptyset \\ 2: & \emptyset \\ 3: & \emptyset \\ 4: & \{(88,99]\}\end{array} 
& \begin{array}{rc} 1: & \{(88,113]\}\\ 2: &  \{(89,113]\}\end{array} \\
\midrule
\Y_{k}(\varphi) :&  & \{(25,30]\} & \{(30,32]\} & \emptyset & \{(88,99]\}\\
\bottomrule
\end{array}
\end{equation*}
\caption{An example run of the network constructed for the formula $\varphi = \psi_1 \sincexq{18}{24} \psi_2$ over dense time behaviors of $\psi_1$ and $\psi_2$.}
\label{tab:run-timed-since-dense}
\end{table}

Table \ref{tab:run-timed-since-dense} illustrates how our sequential network processes a dense time behavior for the formula $\varphi = \psi_1 \sincexq{18}{24} \psi_2$ where $\psi_1$ and $\psi_2$ represent subformulas over dense time. 
Each row corresponds to a step in the processing, progressing through global segments defined by the input behavior and the timing window of the Since operator. 
Within each global segment, the table breaks down the output $\Y(\varphi)$ into local valuations ($||$) for individual time periods. 
These local valuations depend on the satisfaction of both subformulas $\Y(\psi_1)$ and $\Y(\psi_2)$ within the corresponding period and the timing constraints of the Since operation.
The final output $\Y(\varphi)$ for each global segment is obtained by joining the local outputs, reflecting the overall periods where the formula $\varphi$ holds based on the combined behavior of $\psi_1$ and $\psi_2$ across the global segment.

Finally, we establish the correctness of our dense time construction.
Our strategy involves two steps. 
First, we reduce the dense-time behavior to an equivalent, evenly-segmented form compatible with the timing constraints specified in the formula. 
Then, we prove the correctness of our procedure over this reduced behavior, without loss of generality. 
The structure of the dense-time proof deliberately is similar to the discrete case.

\begin{prop}
Given a \PastMTL formula $\varphi$, there exists a base of quantization for any dense time behavior $w$ is closed under dense time sequential network update functions.
\label{thm:dense-quanta}
\end{prop}
\begin{proof}
Let $\delta\in\mathbb{Q}^{+}$ be the greatest rational common divisor of all segment boundaries in the behavior $w$ and timing constraints in the formula $\varphi$. The base of quantization generated by~$\delta$ is trivially closed under addition; thus dense time Since update function in Equation~\ref{DefUpdSinceDense} does not generate new segment boundaries outside the existing base. Since Boolean update functions trivially preserve the existing base, the proof is complete.
\end{proof}

\begin{thm}
For any \PastMTL formula $\varphi$ and dense time behavior $w$, the sequential network output $\Y_k(\varphi)$ evaluates to true at time point $k$ over $w$ iff $(w, k) \vDash \varphi$ holds.
\label{thm:dense-correctness}
\end{thm}
\begin{proof}
Let an evenly segmented behavior $w'$ that is equivalent to the behavior $w$ and compatible with the formula $\varphi$ by Proposition~\ref{thm:dense-quanta}. We assume the fixed segment length $\delta=1$ without loss of generality and proceed similarly to the proof of discrete time Theorem~\ref{thm:discrete-correctness}. Boolean operations are straightforward. The case for dense time Since is as follows: 
\begin{enumerate}
\item[($\shortrightarrow$)] First, we show that $(w, t) \vDash \psi_{1}\sincex{a}{b}\psi_{2}$ holds for all $t \in (k, k+1]$ if a segment $(k, k+1] \subseteq \V_{k}(\psi_{1}\sincex{a}{b}\psi_{2})$. Assume that $(k, k+1] \subseteq \V_{k}(\psi_{1}\sincex{a}{b}\psi_{2})$. By Definition~\ref{DefSd} of the dense time update function, at least one of the following cases holds: 
\begin{enumerate}[(i)]
\item The segment $(k, k+1]$ was added to the valuation set by a segment $(k^{\dagger}, k^{\dagger}+1] \subseteq (k-b, k-a]$ where $\varphi_{2}$ and $\neg\varphi_{1}$ holds. Furthermore, the formula $\varphi_1$ holds for all points in $(k^{\dagger}+1, k+1]$. Then, the intersection of $(k-b, k-a]$ and $[t-b, t-a)$ for all $t \in (k, k+1]$ is non-empty. Hence, $(w, t) \vDash \psi_{1}$ for all $t \in (k, k+1]$.
\item The segment $(k, k+1]$ was added to the valuation set by a segment $(k^{\dagger}, k^{\dagger}+1] \subseteq (k-b, k-a+1]$ where $\varphi_{2}$ and $\varphi_{1}$ holds. Furthermore, the formula $\varphi_1$ holds for all points in $(k^{\dagger}, k+1]$. Then, the intersection of $(k-b, k-a+1]$ and $[t-b, t-a)$ for all $t \in (k, k+1]$ is non-empty.
Hence, $(w, t) \vDash \psi_{1}$ for all $t \in (k, k+1]$.
\end{enumerate}
The semantic definition of the Since operation is satisfied in either case, and we conclude $(w, t) \vDash \psi_{1}$ for all $t \in (k, k+1]$.

\item[($\shortleftarrow$)] Second, we show that $(w, t) \nvDash \psi_{1}\sincex{a}{b}\psi_{2}$ holds for all $t \in (k, k+1]$ if a segment $(k, k+1] \nsubseteq \V_{k}(\psi_{1}\sincex{a}{b}\psi_{2})$. 
Assume that $(k, k+1] \nsubseteq \V_{k}(\psi_{1}\sincex{a}{b}\psi_{2})$. There are two possibilities:
\begin{enumerate}[(i)]
\item The segment $(k, k+1]$ is never added to $\V_{k}$. Then, there exists no segment that satisfies $\psi_{2}$ and timing constraints, thus $(w, k) \nvDash \psi_{1}\sincex{a}{b}\psi_{2}$.
\item The segment $(k, k+1]$ is initially added to $\V_{k}$ at some segment $k^{\dagger}$ and removed at a later step. This means there exists a segment $k^{\ddagger}$ between $k^{\dagger} < k^{\ddagger} \leq k$ where $\psi_{1}$ does not hold as the update function only removes time points from $\V_{k}$ when $\psi_{1}$ does not hold. Thus, $(w, k) \nvDash \psi_{1}\sincex{a}{b}\psi_{2}$.
\end{enumerate}
The semantic definition is violated in either case, and we conclude $(w, k) \nvDash \psi_{1}\sincex{a}{b}\psi_{2}$.
\end{enumerate}
This completes both directions of the equivalence proof, establishing the correctness of the dense time update function for the timed Since operator and our construction.
\end{proof}

\noindent The quantization technique employed in the proof is not practical, however. 
In practice, the number of quantized segments may be very large and sensitive to the specific numeric constants present in the behavior and timing constraints.
This would bring a higher and unnecessary computational burden in many dense time monitoring applications. 
Therefore, our dense time procedure performs a lazy and local synchronization on the fly, only dividing time intervals as needed during the evaluation process.

\section{Implementation and Evaluation}
\label{sec:eval}

This section presents a comparative performance and scalability analysis of the Reelay monitoring library that implements sequential network-based monitors as explained in this paper. 
Reelay\footnote{\url{https://github.com/doganulus/reelay}} provides a comprehensive solution for specification-based monitoring of temporal behaviors, offering flexibility and extensibility at its core.
Its latest version is written as a header-only C++ template library.
Reelay implements sequential networks as an acyclic computation graph using runtime polymorphism in C++, which provides a flexible structure and easier usage but incurs virtual function indirection overhead.
The software design permits users to customize the types for input streams, allowing them to opt for their preferred representation formats for temporal behaviors, provided they create the necessary adapter code.
Furthermore, as a standalone monitoring tool, Reelay releases include a C++ application \texttt{ryjson}, which leverages the \texttt{reelay} library to monitor the newline-delimited sequences of JSON documents, which represents discrete and dense time behaviors.
This application is exclusively utilized for all performance experiments conducted in this study.

We start our analysis with some basic experiments to demonstrate the benefits and limitations of the MTL monitoring technique explained in this paper. 
In particular, we compare existing techniques implemented in monitoring tools against certain challenging properties on synthetic and adversarial inputs. 
This analysis clarifies the importance and effectiveness of our approach in tackling large timing constraint issues, especially when compared to contemporary state-of-the-art solutions. 
In this part, we also compare discrete and dense time monitoring, delving into their strengths and weaknesses within various monitoring contexts based on performance results. 

The second and third parts of the analysis demonstrate the typical performance of discrete and dense time monitors over Timescales benchmarks~\cite{ulus2019timescales}. 
Discrete time benchmarks compare Reelay with other publicly available MTL monitoring tools, whereas dense time benchmarks are compared with the discrete case.
In all our experiments, we report performance results from our Linux-based containerized benchmarking environment, which runs on a six-core 3.80GHz Intel Xeon W-2235 CPU. 
Benchmark scripts and repeatability instructions are available in our code repository\footnote{\url{https://github.com/doganulus/timescales}}.

\subsection{Basic Experiments}
\label{sec:eval-part1}
As detailed in Sections~\ref{sec:discrete} and~\ref{sec:dense}, our monitor construction technique generates a set of nodes with update and output functions based on the syntax tree of the \PastMTL formula.
These nodes form an acyclic computation graph where each performs certain Boolean and interval set operations depending on the node type and the time model at every computation step.
This section compares our approach with other alternative approaches implemented in two publicly available monitoring tools, MonPoly~\cite{monpoly,monpoly2} and Aerial~\cite{aerial},  that support online \PastMTL monitoring.
Both tools use a sample-based time model, equivalent to the discrete model if there is a sample for each discrete time point.
Conversely, their sample-based model and our interval-based model for dense time behaviors are inherently different.
Therefore, we restrict our evaluation to discrete time behaviors only for a fair comparison.

To systematically evaluate the scalability of timed monitoring approaches under different temporal constraints and behavior lengths, we conduct a number of experiments using three parameterized \PastMTL properties.
We define three variants for each property, each encompassing a tenfold increase in timing constraints, spanning from ten to thousand time units.
These properties are tested over temporal behaviors whose lengths range from ten thousand to one million.
Ideally, we expect the performance of monitoring tools to scale linearly to the behavior length while exhibiting constant execution times with respect to varying timing constraints---a particularly important metric for real-time systems where specified timing constraints can be significantly larger than the system's base time unit.

\begin{table}[b]
\centering
\begin{tabular}{
lrrrcrrrcrrr
}
\toprule
& \multicolumn{3}{c}{Aerial} &\qquad & \multicolumn{3}{c}{MonPoly} &\qquad & \multicolumn{3}{c}{Reelay}\\
& 10K & 100K & 1000K  & & 10K & 100K & 1000K & &10K & 100K & 1000K \\
\midrule
\texttt{QPR10} & 0.031 & 0.303 & 3.413 & & 0.025 & 0.285 & 3.062 & &\bfseries 0.020 &\bfseries 0.037 &\bfseries 0.255 \\
\texttt{QPR100} & 0.270 & 2.658 & 27.418 & & 0.024 & 0.274 & 2.861 & &\bfseries 0.020 &\bfseries 0.036 &\bfseries 0.263\\
\texttt{QPR1000} & 3.859 & 38.345 & 315.566 & & 0.026 & 0.275 & 2.911 & &\bfseries 0.020 &\bfseries 0.036 &\bfseries 0.260\\
\midrule
\texttt{PandQ10} & 0.038 & 0.267 & 2.544 & & 0.032 & 0.349 & 3.535 & &\bfseries 0.021 &\bfseries 0.046 &\bfseries 0.367\\
\texttt{PandQ100} & 0.243 & 2.275 & 22.723 & & 0.066 & 0.639 & 6.500 & &\bfseries 0.021 &\bfseries 0.055 &\bfseries 0.371\\
\texttt{PandQ1000} & 3.215 & 32.039 & 401.784 & & 0.339 & 3.664 & 36.786 & &\bfseries 0.021 &\bfseries 0.048 &\bfseries 0.365\\
\midrule
\texttt{Delay10} & 0.039 & 0.252 & 2.538 & & 0.028 & 0.300 & 3.048 & &\bfseries 0.021 &\bfseries 0.046 &\bfseries 0.387 \\
\texttt{Delay100} & 0.215 & 2.126 & 21.015 & & 0.036 & 0.404 & 4.243 & &\bfseries 0.026 &\bfseries 0.133 &\bfseries 1.280 \\
\texttt{Delay1000} & 2.952 & 29.339 & 294.031 & & 0.142 & 1.551 & 15.304 & &\bfseries 0.087 &\bfseries 1.100 &\bfseries 10.523\\
\bottomrule
\end{tabular}
\caption{Total execution times in seconds for discrete time monitoring tools over three different properties and behavior lengths.}
\label{tab:result-discrete}
\end{table}

Table \ref{tab:result-discrete} presents our evaluation results for these performance experiments.
First, the property $\texttt{QPR}: \hist\big((r \wedge \neg q \wedge\once q) \rightarrow (p\ \sincex{l}{u} q)\big)$ is a typical \PastMTL property where we consider three variants, \texttt{QPR10}, \texttt{QPR100}, and \texttt{QPR1000}, whose timing parameters are defined as [3:6], [30:60], and  [300:600], respectively.
We see that MonPoly and Reelay handle longer temporal behaviors and large timing constraints as expected for \texttt{QPR} property. 
Aerial also handles longer behaviors as expected. 
Yet it does not scale for large timing constraints as it uses naive discretization for handling timing constraints.
For these properties and behaviors, Reelay appears 10-15 times faster than MonPoly, yet the performance difference is still in the margin of the implementation details between tools ---especially regarding the programming language choices of C++ (Reelay) and OCaml (MonPoly, Aerial).

Second, the property $\texttt{PandQ}: p\ \sincex{a}{b} q$ is a basic property where we consider three variants, \texttt{PandQ10}, \texttt{PandQ100}, and \texttt{PandQ1000}, whose timing parameters are defined as [1:6], [1:60], and  [1:600], respectively.
We evaluate these three variants over a specific temporal behavior where the proposition $q$ occurs frequently.
This combination attacks MonPoly's approach of keeping bounded history windows for timed nodes.
The performance results in Table \ref{tab:result-discrete} show that MonPoly does not scale for this particular case as MonPoly explicitly stores all occurrences of the second argument of Since operation in the history window bounded by the constraint $b$.
For Reelay, frequent occurrence of $q$ leads to overlapping marked periods in the future that we can merge, thus keeping the size of the valuation set small.
Therefore, our approach performs well in this case and requires a constant time as the timing constraint $b$ gets larger.

However, for the third property, we attack our own method by a very specific adversarial case that involves (1) a formula $\oncex{a}{b} q$ with a very precise and large timing constraint such that $b-a \ll b$, and (2) an input behavior of the proposition $q$ holds very frequently such as on every other time point.
This pathological scenario is captured by the property $\texttt{Delay}$ where we consider three variants, \texttt{Delay10}, \texttt{Delay100}, and \texttt{Delay1000}, whose timing parameters are defined as [6:6], [60:60], and  [600:600], respectively.
The performance results in Table \ref{tab:result-discrete} show that the  $\texttt{Delay}$ property does not scale for all three tools.
For Reelay, this case leads to linear growth of the valuation set with respect to the constraint $b$, thus deteriorating the performance as $b$ grows larger.
We present this case as a pathological example of timed monitoring applications.

\begin{table}[p]
\renewcommand\tabularxcolumn[1]{>{\RaggedRight}m{#1}}
\scriptsize
\centering
\begin{tabularx}{\textwidth}{>{\small}Xcl>{\tiny}c}
\toprule
\normalsize Timescales Properties & \quad &\normalsize \MTL Formulas\\
\midrule
\textbf{Absence After Q}\newline
Always the case that the proposition $p$ does not occur at least for \texttt{b} time units after the proposition $q$ occurs. 
& &
$\begin{array}{l}\hist \big(\oncex{}{b}\ q \longrightarrow (\neg p\since q)\big)\end{array}$
&
(\texttt{AbsentAQ})
\\
\midrule
\textbf{Absence Before R}\newline
Always the case that the proposition $p$ does not occur at least for \texttt{b} time units before the proposition $r$ occurs. 
& &
$\begin{array}{l}\hist \big(r \longrightarrow \histx{}{b}\neg p \big)\end{array}$
&
(\texttt{AbsentBR})
\\
\midrule
\textbf{Absence Between Q and R}\newline
Always the case that the proposition $p$ does not occur between propositions $q$ and $r$ and the duration between $q$ and $r$ is in \texttt{a} and \texttt{b} time units.
& &
$\begin{array}{l}\hist \big((r\wedge \neg q \wedge \once q) \longrightarrow (\neg p\ \sincex{a}{b}\ q)\big)\end{array}$
&
(\texttt{AbsentBQR})
\\
\midrule
\textbf{Universality After Q}\newline
Always the case that the proposition $p$ always occurs at least for \texttt{b} time units after the proposition $q$ occurs
& &
$\begin{array}{l}\hist \big(\oncex{}{b}\ q \longrightarrow (p\since q)\big)\end{array}$
&
(\texttt{AlwaysAQ})
\\
\midrule
\textbf{Universality Before R}\newline
Always the case that the proposition $p$ always occurs at least for \texttt{b} time units before the proposition $r$ occurs. 
& &
$\begin{array}{l}\hist \big( r \longrightarrow \histx{}{b} p \big)\end{array}$
&
(\texttt{AlwaysBR})
\\
\midrule
\textbf{\mbox{Universality Between Q and R}}\newline
Always the case that the proposition $p$ always occurs between propositions $q$ and $r$, and the duration between $q$ and $r$ is in \texttt{a} and \texttt{b} time units. 
& &
$\begin{array}{l}\hist \big((r\wedge \neg q \wedge \once q) \longrightarrow (p\ \sincex{a}{b}\ q)\big)\end{array}$
&
(\texttt{AlwaysBQR})
\\
\midrule
\textbf{Recurrence Globally}\newline
Always the case that the proposition $p$ occurs at least for every \texttt{b} time unit. 
& &
$\begin{array}{l}
\hist \big(\oncex{}{b} p \big)
\end{array}$
&
(\texttt{RecurGLB})
\\
\midrule
\textbf{Recurrence Between Q and R}\newline
Always the case that the proposition $p$ occurs at least for every \texttt{b} time unit between propositions $q$ and $r$. 
& &
$\begin{array}{l}\hist \Big((r\wedge \neg q \wedge \once q) \longrightarrow
(\oncex{}{b}(p \vee q)\since q)\Big)\end{array}$
&
(\texttt{RecurBQR})
\\
\midrule
\textbf{Response Globally}\newline
Always the case that the proposition $s$ responds to the proposition $p$ in \texttt{a} and \texttt{b} time units. 
& &
$\begin{array}{l}\hist \big((s \longrightarrow \oncex{a}{b} p) \wedge \neg( \neg s\ \sincex{b}{}\ p )\big)\end{array}$
&
(\texttt{RespondGLB})
\\
\midrule
\textbf{Response Between Q and R.}\newline
Always the case that the proposition $s$ responds to the proposition $p$ in \texttt{a} and \texttt{b} time units between propositions $q$ and $r$.
& &
$\begin{array}{l}
\hist \Big((r\wedge \neg q \wedge \once q) \longrightarrow
\\ \quad\quad\quad
\big((s \longrightarrow \oncex{a}{b} p) \wedge \neg( \neg s\ \sincex{b}{} p )\big)\Big)
\end{array}$
&
(\texttt{RespondBQR})
\\
\bottomrule
\end{tabularx}
\caption{Parameterized Timescales Properties~\cite{ulus2019timescales}}
\label{tab:properties}
\end{table}

\begin{table}[t]
\centering
\begin{tabular}{>{\ttfamily\footnotesize}lrrr|>{\ttfamily\footnotesize}lrrr}
\toprule
& Aerial & MonPoly & Reelay && Aerial & MonPoly & Reelay\\
\midrule
AbsentAQ10     & 3.277   & 6.179 &\bfseries 0.396 &
AlwaysAQ10     & 3.347   & 6.241 &\bfseries 0.381 \\
AbsentAQ100    & 15.378  & 6.141 &\bfseries 0.369 &
AlwaysAQ100    & 15.210  & 6.234 &\bfseries 0.350 \\
AbsentAQ1000   & 166.553 & 6.200 &\bfseries 0.366 &
AlwaysAQ1000   & 166.587 & 6.188 &\bfseries 0.349 \\
\midrule
AbsentBQR10    & 3.768   & 7.513 &\bfseries 0.529 &
AlwaysBR10     & 3.332   & 5.910 &\bfseries 0.408 \\
AbsentBQR100   & 17.987  & 7.544 &\bfseries 0.487 &
AlwaysBR100    & 17.702  & 5.968 &\bfseries 0.417 \\
AbsentBQR1000  & 191.937 & 7.327 &\bfseries 0.484 &
AlwaysBR1000   & 193.257 & 5.998 &\bfseries 0.412 \\
\midrule
AbsentBR10     & 2.979   & 5.851 &\bfseries 0.422 & 
AlwaysBQR10    & 3.673   & 7.659 &\bfseries 0.511 \\
AbsentBR100    & 14.901  & 5.761 &\bfseries 0.420 & 
AlwaysBQR100   & 15.479  & 7.688 &\bfseries 0.473 \\
AbsentBR1000   & 166.440 & 5.784 &\bfseries 0.421 & 
AlwaysBQR1000  & 166.743 & 7.646 &\bfseries 0.466 \\
\midrule
RecurGLB10     & 2.867   & 4.958 &\bfseries 0.304 & 
RecurBQR10     & 4.274   & 7.890 &\bfseries 0.626 \\
RecurGLB100    & 14.525  & 4.790 &\bfseries 0.233 & 
RecurBQR100    & 20.961  & 7.769 &\bfseries 0.565 \\
RecurGLB1000   & 164.242 & 4.826 &\bfseries 0.227 &
RecurBQR1000   & 220.845 & 7.753 &\bfseries 0.549 \\
\midrule
RespondGLB10   & 4.968   & 6.368 &\bfseries 0.566 & 
RespondBQR10   & 5.709   & 9.525 &\bfseries 0.844 \\
RespondGLB100  & 31.062  & 6.255 &\bfseries 0.463 & 
RespondBQR100  & 31.797  & 9.586 &\bfseries 0.749 \\
RespondGLB1000 & 356.832 & 6.178 &\bfseries 0.450 & 
RespondBQR1000 & 363.070 & 9.481 &\bfseries 0.748 \\
\bottomrule
\end{tabular}
\caption{Total execution times in seconds for discrete time monitoring tools over Timescales benchmarks}
\label{tab:table_verif_perf_discrete}
\end{table}

\subsection{Discrete Time Benchmarks}
\label{sec:eval-discrete}
This section presents a comparative analysis of performance benchmarks for our discrete time monitors against existing publicly available tools. 
To provide a rigorous evaluation, we leverage the Timescales benchmark generator~\cite{ulus2019timescales}, enabling the generation of semi-randomized temporal behaviors satisfying predefined temporal logic formulas at every time point.
Table~\ref{tab:properties} details the 10 Timescales properties used in our experiments, representing real-world scenarios and as reliable proxies for assessing typical monitoring tool performance.
Following the methodology from the previous section, we evaluate each property through three variants featuring a tenfold difference in their timing parameters.
These benchmarks demonstrate the performance and scalability of monitoring tools over typical scenarios, complementing our findings in Section~\ref{sec:eval-part1}. 

Table~\ref{tab:table_verif_perf_discrete} presents a performance comparison of the monitoring tools (Aerial, MonPoly, Reelay) across Timescales properties over discrete time behaviors with a length of 1 million.
These results reinforce the evidence of the practical scalability of Reelay and MonPoly in the face of varying timing constraints, as both maintain near-constant execution times over many different properties.
Conversely, Aerial demonstrates its limitation in handling large constraints and does not scale at all.
Regarding absolute speed, Reelay's performance is consistently faster than MonPoly in these benchmarks.
As the speed of executions is very important for many online and offline applications, we must ensure monitoring tools are optimized for speed.
Therefore, we strongly advocate for compiled and system languages in developing monitoring tools, ensuring they can effectively meet the stringent performance demands of modern applications.

\subsection{Dense Time Benchmarks}
\label{sec:eval-dense}

\begin{table}[b]
\centering
\begin{tabular}{>{\ttfamily\footnotesize}lrrcrrcrrcrr}
\toprule
& \ttfamily Discrete &  \ttfamily Dense1 &\quad & \multicolumn{2}{c}{\ttfamily Dense10} &\quad& \multicolumn{2}{c}{\ttfamily Dense100} &\quad& \multicolumn{2}{c}{\ttfamily Dense}\\
& Time & Time &\quad & Len & Time &\quad& Len & Time &\quad& Len & Time\\
\midrule
\ttfamily\footnotesize AbsentAQ10 &\bfseries 0.168 & 0.767 && 345K & 0.276 && 333K & 0.269 && 333K & 0.269\\
\ttfamily\footnotesize AbsentAQ100 &\bfseries 0.157 & 0.731 && 303K & 0.219 && 259K & 0.185 && 258K & 0.183\\
\ttfamily\footnotesize AbsentAQ1000 &\bfseries 0.154 & 0.731 && 296K & 0.211 && 255K & 0.178 && 250K & 0.176\\
\ttfamily\footnotesize AbsentBQR10 &\bfseries 0.264 & 0.743 && 444K & 0.386 && 444K & 0.386 && 444K & 0.385\\
\ttfamily\footnotesize AbsentBQR100 & 0.242 & 0.668 && 137K & 0.120 && 59K & 0.069 && 59K &\bfseries 0.068\\
\ttfamily\footnotesize AbsentBQR1000 & 0.239 & 0.657 && 95K & 0.081 && 15K & 0.029 && 6K &\bfseries 0.023\\
\ttfamily\footnotesize AbsentBR10 &\bfseries 0.181 & 0.607 && 345K & 0.243 && 333K & 0.236 && 333K & 0.236\\
\ttfamily\footnotesize AbsentBR100 &\bfseries 0.180 & 0.604 && 303K & 0.210 && 260K & 0.186 && 258K & 0.184\\
\ttfamily\footnotesize AbsentBR1000 &\bfseries 0.181 & 0.603 && 296K & 0.206 && 254K & 0.183 && 251K &\bfseries 0.181\\
\ttfamily\footnotesize AlwaysAQ10 &\bfseries 0.168 & 0.804 && 345K & 0.279 && 333K & 0.268 && 333K & 0.267\\
\ttfamily\footnotesize AlwaysAQ100 &\bfseries 0.159 & 0.761 && 303K & 0.215 && 260K & 0.174 && 258K & 0.173\\
\ttfamily\footnotesize AlwaysAQ1000 &\bfseries 0.156 & 0.758 && 295K & 0.205 && 255K & 0.167 && 250K & 0.163\\
\ttfamily\footnotesize AlwaysBQR10 &\bfseries 0.248 & 1.033 && 444K & 0.514 && 444K & 0.512 && 444K & 0.512\\
\ttfamily\footnotesize AlwaysBQR100 & 0.232 & 0.949 && 137K & 0.159 && 59K & 0.087 && 59K &\bfseries 0.087\\
\ttfamily\footnotesize AlwaysBQR1000 & 0.230 & 0.943 && 95K & 0.109 && 15K & 0.034 && 6K &\bfseries 0.025\\
\ttfamily\footnotesize AlwaysBR10 &\bfseries 0.180 & 0.642 && 345K & 0.239 && 333K & 0.231 && 332K & 0.232\\
\ttfamily\footnotesize AlwaysBR100 & 0.180 & 0.636 && 303K & 0.204 && 260K & 0.176 && 258K &\bfseries 0.175\\
\ttfamily\footnotesize AlwaysBR1000 & 0.179 & 0.634 && 296K & 0.197 && 255K & 0.172 && 251K &\bfseries 0.169\\
\ttfamily\footnotesize RecurGLB10 &\bfseries 0.137 & 0.404 && 327K & 0.166 && 327K & 0.167 && 327K & 0.166\\
\ttfamily\footnotesize RecurGLB100 & 0.106 & 0.362 && 117K & 0.063 && 39K & 0.037 && 39K &\bfseries 0.036\\
\ttfamily\footnotesize RecurGLB1000 & 0.104 & 0.359 && 93K & 0.051 && 12K & 0.023 && 6K &\bfseries 0.020\\
\ttfamily\footnotesize RecurBQR10 &\bfseries 0.303 & 1.067 && 336K & 0.426 && 336K & 0.421 && 335K & 0.421\\
\ttfamily\footnotesize RecurBQR100 & 0.278 & 0.993 && 117K & 0.142 && 39K & 0.066 && 39K &\bfseries 0.65\\
\ttfamily\footnotesize RecurBQR1000 & 0.276 & 0.991 && 93K & 0.111 && 12K & 0.032 && 4K &\bfseries 0.023\\
\ttfamily\footnotesize RespondGLB10 &\bfseries 0.262 & 0.974 && 375K & 0.454 && 374K & 0.451 && 374K & 0.452\\
\ttfamily\footnotesize RespondGLB100 & 0.220 & 0.866 && 124K & 0.139 && 45K & 0.072 && 44K &\bfseries 0.072\\
\ttfamily\footnotesize RespondGLB1000 & 0.217 & 0.863 && 94K & 0.100 && 13K & 0.032 && 4K &\bfseries 0.023\\
\ttfamily\footnotesize RespondBQR10 &\bfseries 0.414 & 1.591 && 487K & 0.863 && 487K & 0.862 && 487K & 0.859\\
\ttfamily\footnotesize RespondBQR100 & 0.379 & 1.492 && 147K & 0.253 && 69K & 0.142 && 69K &\bfseries 0.141\\
\ttfamily\footnotesize RespondBQR1000 & 0.383 & 1.485 && 97K & 0.161 && 16K & 0.045 && 7K &\bfseries 0.031\\
\bottomrule
\end{tabular}
\caption{Total execution times in seconds for discrete and dense time monitoring over Timescales benchmarks using Reelay monitoring tool}
\label{tab:table_verif_perf_dense}
\end{table}

In this section, we present a comprehensive performance analysis and benchmarking of our dense time monitor implementations.
We use the same set of Timescales properties and discrete time behaviors defined in Section~\ref{sec:eval-discrete} for dense time experiments after \emph{condensing} these behaviors, to enable a direct comparison of time models and their trade-offs.
Recall that we interpret the discrete time point $k$ with a value $v$ as a constant segment of $(k,k+1]$ with the value $v$ under the dense time setting. 
To enrich our experiments, we consider an additional parameter that controls the maximum duration for condensed periods. 
This parameter is particularly helpful for simulating real-time systems that must respond at a minimum rate. 
In the extreme case, setting the parameter to 1 disables condensation, resulting in the original discrete time representation. 
Conversely, setting it to a sufficiently large value eliminates all stuttering periods, leading to the most condensed representation. 
It is easy to see that condensation is a lossless compression operation for dense time behaviors.

Table~\ref{tab:table_verif_perf_dense} presents the performance of our dense time monitors on dense time Timescales benchmarks. 
These benchmarks were generated using different values for the condensation parameter, as explained earlier. 
The benchmark sets, named \texttt{Dense1}, \texttt{Dense10}, and \texttt{Dense100}, represent scenarios where the maximum duration of a condensed segment is restricted to 1, 10, and 100 time units, respectively. 
The \texttt{Dense} benchmark set represents the scenario with no stuttering periods in input behaviors.
Len columns in Table~\ref{tab:table_verif_perf_dense} indicate the length of the condensed behavior, which originally had a length of 1 million units.
Hence, condensation achieves a compression ratio between 56\% and 99\% over this particular set of behaviors.
Note that this is an important practical advantage for the dense time setting in reducing log storage requirements significantly.

First, we present execution times for our dense time monitors for each property using uncondensed (discrete) input behaviors under Column \texttt{Dense1}. 
For comparison, the discrete time performance for the same inputs is shown under Column \texttt{Discrete}. 
As expected, the discrete time implementation is 3-5 times faster than the dense time implementation due to the absence of synchronization and its use of more primitive data types and fewer memory accesses.
Then, we observe that dense time performance improves as we increase the level of condensation through \texttt{Dense10}, \texttt{Dense100}, and \texttt{Dense} columns. 
This is because our dense time monitor can process condensed behaviors directly and such symbolic treatment corresponds to processing multiple unit intervals at one step.
Consequently, dense time monitors are better suited for hig-frequency temporal behaviors with smaller time steps with fewer changes.
Therefore, dense time performance surpasses discrete time performance for highly condensed behaviors, demonstrating the effectiveness of our approach.

\section{Related Work}
\label{sec:related}
A pioneering work in temporal logic monitoring is the work by Moszkowski to specify and interpret properties of digital circuits and other discrete systems in~\cite{moszkowski1984executing}.
Following his separation theorem for temporal logic, Gabbay differentiated the role of past and future temporal operators for monitoring applications in~\cite{gabbay1989declarative}.
These early works and separating the role of past and future temporal logic would be critical for developing efficient algorithms for temporal logic monitoring as Havelund and Roşu introduced a simple and effective dynamic programming technique to construct online monitors directly from \PastLTL specifications in~\cite{havelund2004monitoring}.

The previously mentioned works, along with others like~\cite{markey2003model,eisner2003reasoning,haakansson2003generating,finkbeiner2009monitor,bauer2011runtime,monitor-bdd}, laid the groundwork for temporal logic monitoring, but they primarily focused on discrete untimed models.
Real-time systems necessitate precise timing constraints that discrete models struggle to represent accurately. Naive discretization techniques exist, but they introduce significant computational overhead. Additionally, improper discretization can lead to a loss of crucial temporal information, hindering the ability to monitor real-time systems effectively.
Following their successful applications in model checking and formal verification, real-time logics~\cite{koy90,dc,alur1994really} have been employed for specification-based monitoring over two different temporal representations~\cite{basin2018algorithms}.
The first approach models continuous/dense time behaviors as sequences of isolated events (samples) with timestamps and applies to monitor them~\cite{thati2005monitoring, drusinsky2006line, ho2014online, basin2015monitoring}.
While this sample based approach yields computationally tractable monitoring algorithms due to its reliance on discrete time representations, its effectiveness relies on the initial discretization, which may reside outside the control of the monitoring framework. This is a potential source of inaccuracies and inefficiencies.
In contrast, the second branch focuses on timed state sequences~\cite{maler2005real, maler2006mitl, colombo2009safe, baldor2013monitoring}, which are more natural to represent physical phenomena.
Monitoring algorithms for this representation exist on a spectrum of trade-offs between practicality and mathematical rigor. 
Our time model in this paper is compatible with the previous works of~\cite{tkleene,patterns,timed-deriv,compass}, which offers a balanced position on this spectrum without compromising mathematical rigor and practicality.

The monitor construction technique presented  in~\cite{havelund2004monitoring} has been proved very fruitful and extended for timed and quantitative properties in several subsequent works~\cite{ 
reinbacher2014runtime,
dokhanchi2014line,
basin2015monitoring, 
chattopadhyay2020verified,
monitor-bdd,havelund2023tp,mamouras2020online,mamouras2023compositional} and implemented in several monitoring tools~\cite{colombo2009larva, monpoly, aerial, dejavu}.
Other approaches for online \MTL monitoring include automata based~\cite{maler2005real,bensalem2005testing,ho2014online}, tester based~\cite{maler2006mitl}, and the incremental marking procedures~\cite{maler2013monitoring}.
These approaches rely on a point-based continuous-time model compared to our period (segment) based dense time model, as discussed in Section~\ref{sec:dense}.
This is a fundamental theoretical difference, and a comprehensive comparison of point-based and period-based time models can be found in~\cite{benthem1991logic}.
Moreover, no implementation is publicly available or maintained for these automata and tester-based monitors to allow practical comparisons.
The implementation of the incremental algorithm has been reported to work only with an external simulator in~\cite{maler2013monitoring} and is not available in the latest version~\cite{amt2}.
Therefore, implementability remains a significant concern, whereas our network-based technique offer a more practical and easily-implementable solution for dense time monitoring.
Finally, it is worth mentioning that the recent works also consider the monitor verifiability and explainability~\cite{schneider2019formally, chattopadhyay2020verified, basin2022verimon, lima2023explainable}.
The explicit state representation and ease of inspection in sequential network-based monitors could contribute to this line of research.

Synchronous dataflow programming languages, such as Lustre~\cite{lustre} and Esterel~\cite{esterel}, and stream runtime verification frameworks~\cite{d2005lola, pike2010copilot, convent2018tessla, gorostiaga2018striver, faymonville2019streamlab} offer alternative monitoring solutions.
These solutions leverage procedural domain-specific languages (DSLs) to describe temporal properties, contrasting with the declarative logic formulas used in temporal logic monitoring.
While procedural languages hold the advantage of familiarity for many software engineers, existing safety standards often advocate for declarative formal semantics due to their unambiguous nature and verifiability.
This motivates the common practice of embedding temporal logic into procedural monitoring frameworks, with various levels of timing constraints support.
However, existing stream real-time runtime verification solutions are still limited by the aforementioned complications of time-event sequences.
This paper shows that time-state sequences can play nicely with the synchronous paradigm under careful design decisions; hence, it is not an inherent limitation, and our procedures can easily be embedded into these frameworks.

\section{Conclusion}
\label{sec:conc}
The ability to monitor diverse data streams, ranging from high-frequency, evenly sampled data to low-frequency, event-driven data, is equally critical to understanding and analyzing complex engineered systems.
Discrete and dense time models effectively capture these distinct data characteristics. 
Discrete time models excel at handling high-frequency, evenly sampled data, where data points are collected at fixed intervals. 
Conversely, dense time models are better suited for unevenly sampled, variable-frequency data, enabling the capture of physical phenomena and long-term relationships. 
To offer a comprehensive solution, monitoring tools must seamlessly integrate both discrete and dense time models.

This paper presented a temporal logic monitoring solution to handle both time models in a unified manner. 
By leveraging the sequential model of computation, we constructed sequential networks from past metric temporal properties.
Our key technique, future temporal marking, employs a symbolic representation of timelines using interval structures, enabling efficient analysis and implementation. 
Unlike existing approaches, our technique effectively handles the complexities of dense time models by discretizing time into variable-duration symbolic steps. 
This approach clearly delineates our assumptions and distinctions from other methods.
We comprehensively evaluated our methods through extensive benchmarking and comparative analysis across various properties and scenarios.

Future work will explore extending our framework to encompass diverse temporal logic variants, including first-order, robust, and probabilistic extensions. 
This entails generalizing our Boolean algebra-based approach by defining suitable update and output functions for other compatible algebraic structures, along with efficient supporting data structures.
We will continue to make these techniques and libraries openly available to facilitate adoption and community contributions.
Our long term goal is a unified monitoring framework capable of handling a broad spectrum of temporal logic flavors, thereby addressing the fragmentation prevalent in current implementations.

In this paper, we primarily focused on constructing monitors for individual temporal properties. 
However, many real-world applications require the simultaneous monitoring of multiple properties over multiple data streams, which introduces additional complexity and demands advanced optimization techniques. 
To address this challenge, we plan to incorporate well-known compiler optimization techniques such as common subexpression elimination and formula rewriting—into the construction process for multi-property monitors. 
Our compositional approach to monitor construction is particularly well-suited to supporting these optimizations in a modular and scalable manner. 
Ultimately, these enhanced monitoring capabilities, including multi-property handling and optimization, will be seamlessly integrated into our existing cloud-native infrastructure.
This will make our approach more adaptable to real-world deployment scenarios.

\bibliographystyle{alphaurl}
\bibliography{references}

\end{document}